\baselineskip=14pt
\magnification=1200
\vsize=7.6in
\def\gmu{\gamma_\mu}
\def\g5{\gamma_5}
\def\ep2{\epsilon_{\mu\nu}}
\def\h{\textstyle{1\over 2}}
\def\gb{\bar g}
\def\psibar{\bar\psi}
\def\km{K^{(-)}_\mu}
\def\kp{K^{(+)}_\mu}
\def\km5{K^(5 (-)_\mu}
\def\kp5{K^(5 (+)_\mu}
\def\amu{A_{\mu} }
\def\a{\alpha}
\def\b{\beta}
\def\k{\kappa}
\def\kc{\kappa_c} 
\def\l{\lambda}
\def\t{\theta}
\def\r{\rho}
\def\s{\sigma}

\centerline{\bf CHIRAL GAUGE MODELS ON A LATTICE}
\vskip1.0cm
\centerline{\bf Alan Horowitz}
\centerline{Department of Physics, University of Pittsburgh}
\centerline{Pittsburgh, Pennsylvania 15260}
\centerline{email: horowitz@falstaff.phyast.pitt.edu}
\vskip1.0cm

\noindent {\bf Abstract}: Chiral gauge groups acting on a lattice
fermion field are constructed such that all fermion modes
(doublers) have the same charge. Details are given for an abelian
axial gauge group within a perturbative framework. An action based on
this group correctly reproduces the continuum gauge-current anomaly,
while preserving global chiral symmetry, locality, rotational symmetry
and hermiticity. A Wess-Zumino-like scalar field enters naturally to
enforce exact chiral gauge invariance. The degeneracy amongst the
doublers can be lifted as in a vector model.
\vfill
\eject

The construction of satisfactory chiral lattice gauge models has long
been a thorny problem. Consequently, there is as yet no lattice
formulation of electro-weak interactions, even though
we've known how to define vector-like models on lattices for a quarter
century. It was recognized early on [1] and crystallized into theorems
[2] that generic models of free lattice fermions with certain
desirable properties, including chiral symmetry, must have equal
numbers of right- and left-handed fermions with the same quantum
numbers as long as the latter are associated with exactly conserved
charges. This is the famous doubling problem.  We then expect
according to the theorem that left-right asymmetry can be attained if
the fermions couple to a non-conserved current. This accords with
continuum physics.  Indeed in a chiral interaction $j\cdot A$, the 
divergence of the current $j$ will in general be afflicted by anomalies. 
On a lattice, this means that the chiral gauge symmetry must be 
explicitly broken, or equivalently is exact but requires scalar degrees 
of freedom in the action to cancel the dependence on the gauge
parameters. The challenge then is to carefully break the symmetry so
that the anomalies come out correctly without sacrificing sacred cows
like locality, rotational symmetry and hermiticity. This is
what will be done here.

Most approaches to date start with a vector-like model and attempt,
with varying degrees of success and clarity, to remove or decouple
fermionic degrees of freedom to achieve left-right asymmetry [3-7].
References [8,9] give critical reviews of these and other
methods. Perhaps the most promising and scrutinized approach is the
Overlap Method [3], inspired by the 5-dimensional domain-wall fermion
idea [4].  It is fair to say that this and other methods so far are
technically difficult and not easy to analyze.  The very recent
approach by L\"uscher [7] proves the existence of abelian lattice chiral
gauge models with exact gauge invariance when gauge anomalies are
cancelled. If true this is a significant achievement. Though it is not
clear whether this formalism lends itself to practical calculations
and how easily it extends to non-abelian models.

The approach given here contrasts with previous ones in that it is
based on a chiral gauge group, albeit of an unconventional type, and
does not depend on the removal of fermionic degrees of freedom for its
chirality. It treats all doublers identically as in a vector model.
The doublers thus resemble the families ($e$, $\mu$, $\tau$) of the
standard model. One can then deal with the fermion degeneracy the same
way as in a vector model like QCD: a spin-diagonalization to staggered
fermions can be made, or Wilson-like terms can be added.  It seems
possible that the degeneracy can be lifted in such a way that some of
the doublers may actually be interpreted as families.  Smit
constructed models pursuant to this philosophy [10], but there was no
handle on gauge invariance.

The formalism will be illustrated on an abelian axial-vector model.
The gauge group and the action for the fermions are constructed below
as expansions in the coupling constant, though it may be possible to
find closed form, non-perturbative expressions.  The gauge current
anomaly emerges easily as does the gauge-symmetry-restoring
Wess-Zumino-like scalar field familiar from the analysis of anomalous
continuum models [11,12].  In previous lattice approaches (excepting the
L\"uscher formalism) the gauge symmetry is broken more implicitly: the
dependence on the scalar field is not easy to extract.

The main motivation of this work is to lay groundwork for the
eventual formulation of a lattice gauge model consistent with the
phenomenology of electro-weak interactions.  For this purpose it
may not be a drawback if we could not go beyond a perturbative
analysis in the fermion sector. Non-perturbative electro-weak physics
seems confined to the Higgs sector.  The methodology presented here
might lead to insights into how the non-perturbative Higgs sector 
interacts with the fermions. Also, many people suspect that the 
chirality of the $SU(2)\times U(1)$ coupling is fundamentally related to 
the presence of scalar fields, the symmetry breaking, and the massiveness
of the vector bosons.  Perhaps a lattice approach, such as the one here,
can vindicate this suspicion.

To simplify expressions I will use the following notation:
  $$\eqalignno{K^{(\pm)}_\mu (m,n) &\equiv \h\{ \psibar_m \gmu \psi_n 
   \pm \psibar_n \gmu \psi_m\} &(1a)\cr
  K^{5(\pm)}_\mu (m,n) &\equiv \h\{\psibar_m \gmu \g5 \psi_n 
  \pm \psibar_n \gmu \g5 \psi_m\}, &(1b)\cr}$$
where $\psi_n$ is a Dirac fermion at lattice site $n$. Units are
chosen whereby the lattice spacing $a=1$. For ease of exposition I
will take the kinetic part of the action to be that of ordinary naive
fermions:
  $$ S_0 = \sum_{n, \mu}K^{(-)}_\mu (n, n+\mu). \eqno(2)$$ 
As is well known, this action contains $2^d$ flavors of fermions
corresponding to the poles in the propagator at the corners of the
Brillouin zone. This degeneracy is a consequence of the invariance of
$S_0$ under ``doubling" transformations, $\psi_n\rightarrow
T\,\psi_n$, where $T = (-1)^{n_\mu}\gmu\g5$ or any product of these
[13,1].  Including the identity, this is a set of $2^d$ elements,
denoted by $T^{(n)}$.  If $\psi^{(0)}_n= u(p)\exp(ipn)$, with
$p\approx 0$, satisfies the continuum Dirac equation, then so do the
doublers $\psi^{(n)}=T^{(n)}\psi^{(0)}$.

A naive attempt to construct a chiral model, by gauging the axial
transformation, $\psi_n^\prime = \exp(i g\g5 \t_n)\psi_n$, is well
known to fail. This is most easily seen from doubling
transformations of the interaction, 
$S_1 = ig \sum K^{5(+)}_\mu (m, m+\mu) \amu(m) $. For half of
the $T^{(n)}$, $S_1$ changes sign. Thus half the fermionic
modes have charge $+g$, while the other half have $-g$. It follows
that the model is left-right symmetric. This is not what we
want. The problem can be traced to fact that the naive axial
transformation does not commute with all the $T^{(n)}$, but
only half of them; the rest anticommute. 

One of the central points of this letter is that a unitary, axial
gauge group can be constructed which commutes with all the $T^{(n)}$
and gives rise to a doubler-symmetric action with the expected
continuum behavior. The simplest infinitesimal transformation with the
desired property has $\delta\psi_n$ proportional to terms of the form
$i\t\g5\psi_{n+\s}$, with $\s$=$(\pm 1,\pm 1,\dots)$. There are 2
choices for the parameters $\t$ which respect unitarity, namely
$\t_n+\t_{n+\s}$ and $\t_{n+\s/2}$.  Note that in the second choice,
$\t$ sits on a site of the dual lattice. I will use this choice
because it gives simpler expressions besides being more elegant.  It
also reveals a kind of duality between vector and axial-vector gauge
transformations. (In fact if we omit the $\g5$ from the infinitesimal
transformation to make it vector-like then half the fermions
transforms with positive charge and half with negative, just as in the
naive axial transformation of the previous paragraph.) So to first
order the axial transformation is
  $$ \psi^\prime_n = 
     \psi_n +  i \gb \g5 \sum_\s \t_{n+\s/2} \psi_{n+\s}  \eqno(3) $$
where I have defined $\gb\equiv g/2^d$ for convenience because the 
sum over $\s$ has $2^d$ terms. Although this transformation is a matrix 
in space-time, it commutes with the generators of lattice rotations and
translations.

The higher orders in the axial transformation are determined by
unitarity and the group composition law. In order to form a group it
is necessary that the transformation, starting at $O(g^2)$, involve
the gauge field $\amu$.  (This is reminiscent of, though not directly
related to, Fujikawa's derivation of the anomaly [14], wherein the
fermionic functional measure becomes dependent on $\amu$.)  The group
property can also be ensured by introducing a group-valued scalar
field transforming as $\eta_n^\prime = \eta_n + \t_n$. However, in the
end, the analysis turns out to be simpler using just $\amu$, and
besides, we want to see how far we can go before we have introduce
scalar fields.

Consider 2 successive gauge transformations 
    $$\eqalign{ \psi^\prime &= G[\phi, A] \psi \cr 
      \psi^{\prime\prime}
     &= G[\t, A^\prime] \psi^\prime = G[\t, A + \Delta \phi] G[\phi,
     A] \psi.\cr}$$
In order to have an abelian group we must have
    $$G[\t, A + \Delta \phi] G[\phi, A] = G[\t + \phi, A]. \eqno(4)$$
This condition, along with unitarity, uniquely determines the
$O(g^2)$ term from the $O(g)$ term.  Rewriting the transformation,
Eq. (3), with this term now included:
  $$\eqalignno{\psi^\prime_n = \psi_n + 
     &i \gb \g5 \sum_\s \t_k \psi_{n+\s} &(5)\cr 
   &-\h \gb^2\sum_{\r,\s}\bigl\{ \t_k \t_l - 
      (\t _k+ \t_l)[A_{k,l} + \h (\t_l - \t_k)] \bigr\}
     \psi_{n+\r+\s} + O(g^3) & \cr} $$
where $k \equiv n+\s/2$, $l \equiv n+\s+\r/2$ and $A_{k,l}$ is a sum
over $A$s on dual lattice links forming a path between $k$ and $l$.
The path may be taken to minimize the deviation from
the straight-line path between $k$ and $l$. If there is more than one 
such path, an average is taken.  (In Eq. (5) and below, $\s$ and $\r$
always denote the vectors $(\pm 1,\pm 1,\dots)$, while $\mu$ denotes
a unit vector.) Unlike the $O(g^2)$ term, the $O(g^3)$ term, which I've 
calculated, is not uniquely determined by unitarity and Eq. (4).  There 
are 2 free parameters. 

To go beyond the perturbative analysis of this paper, it would be
necessary to find a unitary, closed form solution of Eq. (4), which 
reduces to the infinitesimal transformation, Eq. (3). I have not yet
made a serious attempt to do this. In such a pursuit, it may or may not
be beneficial to recast Eq. (4) as a system of differential equations
obtained by taking $\phi$ infinitesimal and Taylor expanding. 

It is easy to see that $\det\, G = 1$ to all orders in $g$. This is
done using $\det\, G = \exp({\rm Tr}\, \ln G)$, expanding $\ln\,G$ in
powers of $g$, and noting that all odd powers vanish because ${\rm
Tr}\,\g5=0$ (as does the trace of the space-time matrix at odd
powers). This leaves only even powers which being real must also
vanish because $\det\, G$ can only be a unit phase factor, since $G$
is unitary. Since $\det\, G = 1$, the fermion functional measure is
invariant under the chiral transformation, Eq. (5). The anomaly
therefore does not appear through the variation of the measure: it
appears through the variation of the action.

The local axial gauge group, denoted $\tilde U_1^5$, defined by Eqs. (3), 
(4) and unitarity is not strictly local.  At order $g^N$, $G_{mn}$ has 
non-vanishing elements out to $|n-m| = 2N$ in 4-dimensions. The fermion 
action based on $\tilde U_1^5$ will thus have long range interactions at
higher orders in $g$. However the $O(g^N)$ term in the action,
$\psibar_m\dots\psi_n$, with $max|n-m|\approx 2N$, has dimension
$4+N$, and is thus suppressed by a factor of $a^N$. We therefore
expect no violations of locality in processes with energy far below
the cutoff, $1/a$.

Now I will construct the action through $O(g^2)$ by applying the gauge
principle using $\tilde U_1^5$. Starting from $S_0$ for naive
fermions, Eq. (2), the action, invariant to $O(g)$ under the
transformation, Eq. (5), is $S_0+S_1$ with
 $$ S_1 = -i \gb \sum_{n,\mu,\s} K^{5(+)}_\mu (n, n+\mu+\s/2) 
       \amu (n+\s/2) \eqno(6) $$
where the gauge field $\amu$ sits on links of the dual lattice and has
the usual transformation law: $\amu^\prime = \amu + \Delta_\mu \t$.
In the naive continuum limit, we have the usual axial-vector interaction.

The variation of $S_0 + S_1$ under the transformation can be written
after some algebraic reorganization as
  $$\eqalign{\delta S_{0+1} = \h\gb^2&\sum
    \Bigl\{ K^{(-)}_\mu (n, n+\mu+\r+\s)
 \bigl[(\t_{l+\mu}-\t_k)^2 - \h (\t_l - \t_k)^2 - 
   \h (\t_{l+\mu} - \t_{k+\mu})^2\cr 
  &+2 \t_{l+\mu} \amu(k) -2 \t_k \amu(l) + 
   (\t_{k+\mu} + \t_{l+\mu}) A_{k+\mu,l+\mu} - 
     (\t_k + \t_l) A_{k,l}\bigr] \Bigr\}} $$
where as in Eq. (5), $k \equiv n+\s/2$ and $l \equiv n+\s+\r/2$.  There
is no $S_2$ depending only on $\psi$ and $\amu$ which makes $\delta S$
vanish. The reason is the anomaly. In order to cancel the 
$(\delta \t)^2$ terms we need 
         $$ S_2 = - \h\gb^2 \sum_{n,\mu,\r,\s}\bigl\{ K^{(-)}_\mu
                       (n, n+\mu+\r+\s) [ A_{k,l+\mu}^2 - \h A_{k,l}^2
                       -\h A_{k+\mu,l+\mu}^2 ]\bigr\}. \eqno(7)$$
Now with $S=S_{0+1+2}$ we are left with the anomaly in the form  
  $$\delta S = \gb^2 \sum_{n,\mu,\r,\s} \Bigl\{ K^{(-)}_\mu (n, n+\mu+\r+\s)
    \bigl[\t_k \sum_{P_1} A  + 
     \t _{l+\mu}\sum_{P_2} A \bigr] \Bigr\} \eqno (8)$$
where $P_1$ ($P_2$) is an ordered, closed path with vertices 
$k,\ l+\mu,\ l$ ($k,k+\mu,l+\mu$).  

In order to expose the anomaly in conventional form, we define the
effective action, $W[A]$, by $\exp(-W[A]) = \int D\psibar D\psi
\exp(-S)$.  It follows that $\delta W[A] = W[A+\Delta\t] - W[A] =
-{\rm ln}\langle\exp(-\delta S)\rangle \approx \langle\delta S\rangle$
for small $\t$, the average being over the fermion fields. In both 2 and 4 
dimensions, $\langle\delta S\rangle$ vanishes identically at $O(g^2)$. 
This agrees with the fact that in 2-dimensions there is no gauge anomaly 
in the axial-vector model.

To compute $\langle\delta S\rangle$ in 4-dimensions in the continuum 
limit the following easily derived ingredients are needed: 
     $$ \sum_{P_1} A \approx \sum_{P_2} A \rightarrow  
              {\textstyle{1\over 4}} (\r+\s)_\l F_{\l \mu} $$

    $$\langle K^{(-)}_\mu (n, n+\r+\s) \rangle = 
       2 i g C \epsilon_{\mu\a\b\nu} (\r+\s)_\a F_{\b\nu} + \dots $$

      $$\sum_\s \s_\mu \s_\nu = 16 \delta_{\mu \nu}$$
where 
  $$C = \int_{-\pi}^{\pi} {d^4p \over {(2\pi)^4}} \sin^2p_1 \cos^2p_1
  \prod_{j=2}^4 \cos^4p_j / (\sum \sin^2 p_\l)^2 = {1\over {6\pi^2}}.$$
Then we find the expected result for 16 flavors,
    $$\delta W[A] = -{i g^3 \over {3\pi^2}} \int d^4x
         \t\epsilon_{\mu\a\b\nu}F_{\mu\a}F_{\b\nu} \eqno (9)$$ 
from which it follows that the current $j^5_\mu = \partial W/\partial
\amu$ has the correct anomalous divergence.

Here, $j_\mu^5$ is, of course, the current coupling to the gauge
field, and not the current $J_\mu^5$ associated with the usual global
chiral symmetry $U_A$, [$\psi_n^\prime = \exp(i\g5 \a)\psi_n$].
Without mass or Wilson terms, $J_\mu^5$ remains exactly conserved,
keeping the fermions massless.  As is well known [1], the conservation
of $J_\mu^5$ owes to the cancellation of the $U_A$ anomalies between
the doublers.  So if we extend this chiral fermion method to the
Standard Model, and try to interpret some of the doublers as families,
then we could get a different pattern of $U_A$ anomaly cancellation
than is currently understood in the continuum model, where the
anomalies don't cancel between families. However at this stage this is
not an important consideration, since experimental signatures for the
cancellation pattern, such as baryon number non-conservation (and B-L
conservation) have not been seen, and are not likely to be seen in the
near future.

To make this axial-vector model $exactly$ gauge-invariant it is not
enough to simply add other fermions with appropriate charges to cancel
the anomaly.  This is apparent from Eq. (8): $\langle\delta S\rangle$
will vanish, but $S$ itself is still gauge-variant.  To have exact
chiral gauge-invariance, a group-valued scalar field $\eta_n$ must
be introduced transforming as $\eta_n^\prime = \eta_n + \t_n$. Then
inspection of Eq. (8) shows that
$$\eqalignno{S = S_{0+1+2} -
  \gb^2 &\sum_{n,\mu,\r,\s}\Bigl\{K^{(-)}_\mu (n,n+\mu+\r+\s) 
 \bigl(\eta_k \sum_{P_1} A  + 
  \eta_{l+\mu}\sum_{P_2} A \bigr)\Bigr\} &(10)\cr
 + &\kappa \sum_{n,\mu} \cos[g(\Delta_\mu \eta - \amu)] + O(g^3).& }$$
is gauge-invariant through $O(g^2)$.
I added a gauge-invariant kinetic term for $\eta$, which
is recognized as a fixed-length Higgs term, with $\eta$ being the
phase of the Higgs field. In view of the analysis above, the
linear term in $\eta$ becomes a Wess-Zumino term [15] in the effective
action $W[\eta, \amu]$.  

All mass and degeneracy-lifting terms in this model must be made
gauge-invariant with the use of the scalar field $\eta$: neither the
mass term $\psibar_n\psi_n$ nor the Wilson term
$r\psibar_n\{\psi_{n+\mu} -2\psi_n +\psi_{n-\mu}\}$ is gauge-invariant
by itself: gauge-invariance requires Yukawa interactions as in the
Standard Model.  It is straightforward, using the chiral gauge
transformation, Eq. (5), to construct these interactions. For example,
the gauge-invariant mass term to $O(g^2)$ is
  $$ \psibar_n\psi_n - 2i\gb\sum_\s\psibar_n\g5\psi_{n+\s}\eta_{n+\s/2} 
     -2\gb^2\sum_{\s,\r}\psibar_n\psi_{n+\s+\r}
      \eta_{n+\s/2}\eta_{n+\s+\r/2}\eqno(11)$$ 
which reduces in the naive continuum limit to the expansion of
$\psibar_x\exp{(-2ig\g5\eta_x)}\psi_x$ to $O(g^2)$.  It is interesting
to note that this lattice mass term is not invariant under the $U_A$
global chiral symmetry, while its naive continuum limit is ( provided
$\eta \rightarrow\eta + \t$ under $U_A$).

Because the action is doubler-symmetric, we can reduce the number
of fermion modes by spin-diagonalizing in the usual way [16]: 
a unitary transformation, $\psi_n\rightarrow
\gamma_1^{n_1}\dots\gamma_4^{n_4}\psi_n$, breaks the action 
into 4 identical, independent pieces. Tossing 3 pieces away we have an
action describing 4 Dirac fermions (reduced from 16), with one
fermionic degree of freedom per lattice site (reduced from 4).  Then
terms can be added to lift the mass degeneracy of these fermions.
Interestingly, The interaction $S_1$, linear in $\amu$, in terms of
these staggered fermions is identical to the linear term in Smit's
staggered fermion approach [10] except for the coordinates of the
gauge fields. Higher order terms are however not similar: Smit's
action has no gauge symmetry. 

How should we use the action, Eq. (10) ?  Anomalous models, such as in
Eq. (10) are believed to be nonrenormalizable [see, e.g. ref. 12].  To
obtain a renormalizable continuum model, the gauge anomaly should be
cancelled by other fermions, as for example in a model with 9 fermion
fields, 8 with axial charge, $q=1$, and one with $q=-2$. To extract
physics from such a model, the parameters must be near a critical
point or surface. Ignoring the fermions for the moment, the Higgs
model with just $\eta$ and $\amu$ has a Higgs phase for $\k>\kc$ with
$M_A>0$ and a symmetric phase for $\k<\kc$ with $M_A=0$.  With an eye
towards electro-weak physics, let's say we want to obtain a continuum
limit in the Higgs phase with a 3 parameter model, $\k$, $g$ and $y$,
the coefficient of the mass-Yukawa term in Eq. (11).  At a point in
the Higgs phase near the critical surface, for small $g$ and $y$ the
truncation of the action at some low order in $g$ should be valid. To
calculate some observable of $\psi$ or $\amu$ one could either
perturbatively integrate out the fermions in the path integral and
numerically integrate (simulate) the resulting expression in $\amu$
and $\eta$, or include the fermions in the simulation. One would have
to check that answers were not sensitive to the order of truncation of
the action. 

As we get closer to the critical surface, the fermion mass $m_f$, in
units of inverse lattice spacing, should approach zero (otherwise
there would be no low mass fermion in physical units).  To get
$m_f\rightarrow 0$ we can either tune $y$ to zero as
$\k\rightarrow\kc$ or keep $y$ fixed and rely on the higher orders in
Eq. (11) to reduce $m_f$. At larger values of $y$ we will need
more terms in Eq. (11). Similar reasoning might apply to an
attempt to remove doublers using a Wilson-Yukawa term: the
fluctuations in $\eta$ may conspire to send the effective amplitude of
such a term to zero, giving the doublers a mass much less than the
cut-off. This is what happens in the Smit-Swift model [17,9]. Unlike in
the Smit-Swift model the chirality of the model here does not depend
on the Wilson-Yukawa term. If need be, the number of doublers can be 
reduced by using staggered fermions or other methods.

The extension of the above methodology to a V-A abelian model is
straightforward. We want a transformation, which in the naive
continuum limit reduces to $\psi^\prime(x) = \exp[ig \t_x \h (1-\g5)]
\psi (x)$, and which transforms all doublers the same.  This is
accomplished by the following infinitesimal transformation,
$$\psi^\prime_n = (1+{i\over 2 }g \t_n )\psi_n - {i\over 4 }\gb \g5
 \sum_\r (\t_n + \t_{n+\r} ) \psi_{n+\r}. $$
The higher orders are constructed using the group property, Eq. (4),
and unitarity as was done above for the axial transformation. The 
extension to non-abelian chiral groups is slightly more subtle, 
and will be presented in another publication.

\vskip1cm

\item{[1]} L.H. Karsten and J. Smit, Nucl. Phys. B183, 103 (1981).
\item{[2]} H.B. Nielsen and M. Ninomiya, Nucl. Phys. B185, 20 (1981);
   Nucl. Phys. B193, 173 (1981).
\item{[3]} R. Narayanan and H. Neuberger, Nucl. Phys. B443, 305 (1995).
\item{[4]} D.B. Kaplan, Phys. Lett. B288, 342 (1992).
\item{[5]} A. Borelli, L. Maiani, G.-C. Rossi, R. Sisto and M. Testa, 
  Nucl. Phys. B333, 335 (1990).
\item{[6]} J.L. Alonso, Phys. Rev. D44, 3258 (1991).
\item{[7]} M. L\"uscher, hep-lat/9811032.
\item{[8]} Y. Shamir, Nucl. Phys. (Proc. Suppl.) B47, 212 (1996).
\item{[9]} D.N. Petcher, Nucl. Phys. (Proc. Suppl.) B30, 50 (1993).
\item{[10]} J. Smit, Nucl. Phys. (Proc. Suppl.) B4, 451 (1988); 
  29B,C, 83 (1992).
\item{[11]} E. D'Hoker and E. Fahri, Nucl. Phys. B248, 59 (1984);
    L.D. Fadeev and S.L. Shatashvili, Phys. Lett. B167, 225 (1986).
\item{[12]} R.D. Ball, Phys. Rep. 182, 1 (1989).
\item{[13]} A. Chodos and J.B. Healy, Nucl. Phys. B127, 426 (1977).
\item{[14]} K. Fujikawa, Phys. Rev. Lett. 42, 1195 (1979).
\item{[15]} J. Wess and B. Zumino, Phys. Lett. B37, 95 (1971).
\item{[16]} N. Kawamoto and J. Smit, Nucl. Phys. B192, 100 (1981).
\item{[17]} P. Swift, Phys. Lett. B145, 256 (1984); J. Smit, Acta. Phys.
   Pol. B17, 531 (1986).

\end